\documentclass[conference]{IEEEtran}
\IEEEoverridecommandlockouts
\usepackage[noadjust]{cite}
   
\usepackage{amsmath,amssymb,amsfonts}
\usepackage{algorithmic}
\usepackage{graphicx}
\usepackage{textcomp}
\usepackage{xcolor}
\usepackage{balance}
\usepackage{algorithm2e}
\usepackage{array}
\usepackage{subcaption} 
\usepackage{stfloats}
\usepackage{url}
\usepackage{verbatim}
\usepackage{titlesec}

\def\BibTeX{{\rm B\kern-.05em{\sc i\kern-.025em b}\kern-.08em
    T\kern-.1667em\lower.7ex\hbox{E}\kern-.125emX}}

\DeclareMathOperator*{\argmax}{argmax}

\newcommand{\s}{\mathbf{s}}

\newcommand{\vbf}{\mathbf{v}}

\newcommand{\m}{\mathbf{m}}

\newcommand{\ji}{\jmath}
\DeclareMathOperator{\sinc}{\rm{sinc}}

\DeclareMathOperator{\EX}{\mathbb{E}}

\usepackage{tikz,pgfplots}
\usetikzlibrary{decorations.pathreplacing,plotmarks}
\pgfplotsset{compat=1.7}
\usepackage{textcomp}
\usetikzlibrary{shapes,arrows}
\usetikzlibrary{arrows.meta}
\usepackage{mathtools}
\usepackage{environ}
\makeatletter
\tikzset{%
  block/.style    = {draw, thick, rectangle, minimum height = 3em,
    minimum width = 3em},
  sum/.style      = {draw, circle, node distance = 2cm}, 
    input/.style    = {coordinate}, 
  output/.style   = {coordinate} 
}
\newsavebox{\measure@tikzpicture}
\NewEnviron{scaletikzpicturetowidth}[1]{%
  \def\tikz@width{#1}%
  \begin{lrbox}{\measure@tikzpicture}%
  \BODY
  \end{lrbox}%
  \pgfmathparse{#1/\wd\measure@tikzpicture}%
  \BODY
}
\usetikzlibrary{calc}
\def\centerarc[#1](#2)(#3:#4:#5)
    { \draw[#1] ($(#2)+({#5*cos(#3)},{#5*sin(#3)})$) arc (#3:#4:#5); }

\tikzset{
 thicker/.style={line width=#1\pgflinewidth},
 thicker/.default={2},
}
\def\mic(#1)(#2)(#3);{
    \begin{scope}[shift={(#1)},rotate={#2},scale={#3}]
        \draw (0,0) circle (0.22);
        \draw[thicker=1.5] 
            (-0.22,-0.3) -- (-0.22,0.3);
    \end{scope}
}
\usetikzlibrary{shapes.geometric}
\def\speaker(#1)(#2)(#3);{
    \begin{scope}[shift={(#1)},rotate={#2},scale={#3}]
        \draw (-0.8,-0.40) rectangle (-0.35,0.40);
        \draw[-] (-0.35,0.40) -- (0,0.65) -- (0,-0.65) -- (-0.35,-0.40);
    \end{scope}
}

\tikzset{
 thicker/.style={line width=#1\pgflinewidth},
 thicker/.default={2},
}
\def\mic(#1)(#2)(#3)(#4);{
    \begin{scope}[shift={(#1)},rotate={#2},scale={#3}]
        \draw[#4] (0,0) circle (0.22);
        \draw[thicker=1.5,#4] 
            (-0.22,-0.3) -- (-0.22,0.3);
    \end{scope}
}
\usetikzlibrary{shapes.geometric}
\def\speaker(#1)(#2)(#3);{
    \begin{scope}[shift={(#1)},rotate={#2},scale={#3}]
        \draw (-0.8,-0.40) rectangle (-0.35,0.40);
        \draw[-] (-0.35,0.40) -- (0,0.65) -- (0,-0.65) -- (-0.35,-0.40);
    \end{scope}
}
\usepackage{pifont}

\usetikzlibrary{calc}
\def\centerarc[#1](#2)(#3:#4:#5)
    { \draw[#1] ($(#2)+({#5*cos(#3)},{#5*sin(#3)})$) arc (#3:#4:#5); }
\definecolor{matlabgreen}{rgb}{0.466, 0.674, 0.188}
\definecolor{matlabyellow}{rgb}{0.929, 0.694, 0.125}
\usepackage{pifont,xcolor}
\definecolor{myblue}{RGB}{0,29,119}

\newcounter{MYtempeqncnt} 

\begin{document}

\title{Steered Response Power-Based Direction-of-Arrival Estimation Exploiting an Auxiliary Microphone
\thanks{This work was funded by the Deutsche Forschungsgemeinschaft (DFG, German Research Foundation) under Germany's Excellence Strategy - EXC 2177/1 - Project ID 390895286 and Project ID 352015383 - SFB 1330 B2.
}
}

\author{\IEEEauthorblockN{Klaus Br\"{u}mann and Simon Doclo}
\IEEEauthorblockA{
\textit{Dept. of Medical Physics and Acoustics and Cluster of Excellence Hearing4all,} \\
\textit{Carl von Ossietzky Universit\"{a}t}, Oldenburg, Germany \\
\{klaus.bruemann,simon.doclo\}@uni-oldenburg.de
}
}

\maketitle

\begin{abstract}
Accurately estimating the direction-of-arrival (DOA) of a speech source using a compact microphone array (CMA) is often complicated by background noise and reverberation. A commonly used DOA estimation method is the steered response power with phase transform (SRP-PHAT) function, which has been shown to work reliably in moderate levels of noise and reverberation. Since for closely spaced microphones the spatial coherence of noise and reverberation may be high over an extended frequency range, this may negatively affect the SRP-PHAT spectra, resulting in DOA estimation errors. Assuming the availability of an auxiliary microphone at an unknown position which is spatially separated from the CMA, in this paper we propose to compute the SRP-PHAT spectra between the microphones of the CMA based on the SRP-PHAT spectra between the auxiliary microphone and the microphones of the CMA. For different levels of noise and reverberation, we show how far the auxiliary microphone needs to be spatially separated from the CMA for the auxiliary microphone-based SRP-PHAT spectra to be more reliable than the SRP-PHAT spectra without the auxiliary microphone. These findings are validated based on simulated microphone signals for several auxiliary microphone positions and two different noise and reverberation conditions. 
\end{abstract}

\begin{IEEEkeywords}
Steered-Response Power, Direction-of-arrival, Auxiliary Microphone, Source Localization
\end{IEEEkeywords}

\section{Introduction} 
\label{sec: Introduction} 
In many speech processing applications, compact microphone arrays (CMAs) are used to estimate the direction-of-arrival (DOA) of a speech source. 
A popular DOA estimation method is based on the steered response power (SRP) function \cite{omologo1997use, dibiase2000high, dmochowski2010steered, pertila2018multichannel,dietzen2021low}, where the DOA is estimated as the direction maximizing the SRP spectra over frequencies and microphone pairs. 
Despite its simplicity, the SRP function is rather sensitive to reverberation and spatially coherent noise and has a poor spatial resolution \cite{salvati2014incoherent}. 

To improve robustness against noise and reverberation, several frequency weightings have been proposed \cite{valin2007robust,abutalebi2011performance,athanasopoulos2016contributions, braun2015narrowband}. 
The most common frequency weighting is the  phase transform (PHAT) \cite{knapp1976generalized}, which removes the influence of the amplitude in the SRP spectra, effectively decorrelating the microphone signals and enhancing the spatial resolution. 
While the PHAT weighting is optimal in anechoic environments with spatially uncorrelated noise, it has also been demonstrated to work well in realistic noisy and reverberant acoustic environments where these assumptions are not satisfied \cite{chen2006time,zhang2008does,velasco2016proposal}. 
Nevertheless, for small inter-microphone distances noise and reverberation typically exhibit a larger spatial coherence than for large inter-microphone distances, especially at low frequencies, negatively affecting the SRP-PHAT spectra and the DOA estimation performance. 

In this paper, we assume that an additional microphone is available, which is spatially separated from the CMA. 
The main objective is to exploit this additional microphone to enhance the estimation accuracy of the SRP spectra in noisy and reverberant environments. 
In recent years, several source localization methods exploiting an auxiliary microphone have been proposed. 
While some methods assume that the auxiliary microphone is in the vicinity of the source \cite{farmani2017informed, farmani2018bias, kowalk2022signal}, other methods just assume that the auxiliary microphone is spatially separated from the microphone array \cite{fejgin2021comparison, fejgin2023exploiting, bruemannEmic}. 
In both cases the spatial coherence of the noise and reverberation between the microphones of the array and the auxiliary microphone is typically lower than the spatial coherence between the microphones of the array. 
Similarly to \cite{bruemannEmic}, in this paper we propose to compute the SRP-PHAT spectra between the microphones of the CMA based on the SRP-PHAT spectra between the auxiliary microphone and the microphones of the CMA. 
Assuming a spherically isotropic sound field and a constant power spectrum for the noise and reverberation, we theoretically analyse the influence of the noise and reverberation on the auxiliary microphone-based SRP-PHAT spectrum. 
We show that the auxiliary microphone-based SRP-PHAT spectrum is more reliable than the conventional SRP-PHAT spectrum without the auxiliary microphone when the distance between the auxiliary microphone and the CMA is larger (depending on the noise and reverberation level). 

Based on simulated microphone signals for a speech source in a reverberant room with multi-talker babble noise, we validate these findings for several positions of the auxiliary microphone. 
Simulation results show that for most positions of the auxiliary microphone a notable improvement in terms of DOA estimation accuracy is achieved, even in challenging acoustical conditions. 

\section{SRP-Based DOA Estimation Exploiting an Auxiliary Microphone}
\label{eq: SRP}
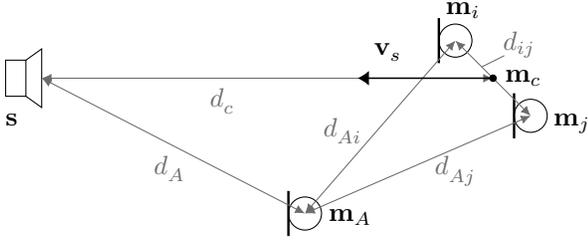
\begin{figure}[t!]
\centering
\begin{tikzpicture}[scale = 1] 
\speaker(14,-11.7)(0)(0.6);
\node[align=center] at (14-0.4,-11.7-0.55) {$\s$};
\mic(19.5,-11.7+0.5)(0)(1)(black); 
\mic(20.5,-11.7-0.5)(0)(1)(black); 
\node[align=center] at (19.5+0.1,-11.7+0.5+0.4) {$\m_{i}$};
\node[align=center] at (20.5+0.55,-11.7-0.5-0.1) {$\m_{j}$};
\draw[gray!85!black,{Latex[length=1.3mm, width=1.2mm]}-{Latex[length=1.3mm, width=1.2mm]}] (14,-11.7) -- (20,-11.7);
\node[gray!85!black,align=center] at (16.4,-11.7-0.3) {$d_{c}^{}$};
\draw[gray!85!black,{Latex[length=1.3mm, width=1.2mm]}-{Latex[length=1.3mm, width=1.2mm]}] (19.5,-11.7+0.5) -- (20.5,-11.7-0.5);
\draw[gray!85!black,-] (20-0.15,-11.7+0.15) -- (20-0.15+0.3,-11.7+0.15+0.15);
\node[gray!85!black,align=center] at (20+0.35,-11.7+0.45) {$d_{ij}$};
\mic(17.5,-13.5)(0)(1)(black); 
\node[align=center] at (18.1,-13.55) {$\m_{A}$};
\draw[gray!85!black,{Latex[length=1.3mm, width=1.2mm]}-{Latex[length=1.3mm, width=1.2mm]}] (14,-11.7) -- (17.5,-13.5);
\node[gray!85!black,align=center] at (15.7,-11.7-1.2) {$d_{A}^{}$};
\draw[gray!85!black,{Latex[length=1.3mm, width=1.2mm]}-{Latex[length=1.3mm, width=1.2mm]}] (17.5,-13.5) -- (19.5,-11.7+0.5);
\node[gray!85!black,align=center] at (18,-11.7-0.7) {$d_{Ai}^{}$};
\draw[gray!85!black,{Latex[length=1.3mm, width=1.2mm]}-{Latex[length=1.3mm, width=1.2mm]}] (17.5,-13.5) -- (20.5,-11.7-0.5);
\node[gray!85!black,align=center] at (19.5,-11.7-1.23) {$d_{Aj}^{}$};
\draw[line width=0.2mm,-{Latex[length=1.5mm, width=2mm]}] (20,-11.7) -- (20-1.8,-11.7);
\node[align=center] at (18.8-0.2,-11.7+0.35) {$\vbf_{s}^{}$}; 
\node at (20,-11.7)[circle,fill,inner sep=1pt]{};
\node[align=center] at (20+0.4,-11.7) {$\m_{c}$};

\end{tikzpicture}
\caption{Exemplary arrangement of a source at position $\s$, closely spaced microphones at positions $\m_{i}$ and $\m_{j}$, and an auxiliary microphone at position $\m_{A}$.\vspace*{-3 mm}}
\label{fig: Scenario}
\end{figure}

We consider an acoustic scenario with a single speech source in a noisy and reverberant environment, recorded with a CMA consisting of $M$ microphones. 
The position of the source is denoted by $\s$ and the position of the $j$-th microphone is denoted by $\m_{j}$ (see Fig. \ref{fig: Scenario}). 

In the continuous-time Fourier transform domain, the reverberant and noisy $j$-th microphone signal $Y_{j}(\omega)$ can be decomposed as \vspace*{-1 mm}
\begin{equation}
Y_{j}(\omega) = X_{j}^{D}(\omega) + \underbrace{X_{j}^{R}(\omega) + N_{j}(\omega)}_{U_{j}(\omega)} \; ,\vspace*{-1 mm}
\end{equation}
where $\omega$ denotes the radial frequency, $X_{j}^{D}(\omega)$ denotes the direct path component and $U_{j}(\omega)$ denotes the undesired component, comprising of the reverberation component $X_{j}^{R}(\omega)$ and the noise component $N_{j}(\omega)$. 
Assuming a free field acoustic scenario, the direct path component is equal to the source signal $S(\omega)$ with a phase shift and attenuation, i.e., \vspace*{-1 mm}
\begin{equation}
X_{j}^{D}(\omega) = \frac{1}{\xi d_{j}} S(\omega) \exp(-\ji \omega d_{j}/\nu) \; ,\label{eq: direct path ATF}\vspace*{-1 mm}
\end{equation}
with $\ji = \sqrt{-1}$, $\xi = {4\pi}$ the attenuation factor of a point source, $d_{j} = ||\s - \m_{j}||$ the distance between the source and the $j$-th microphone, and $\nu$ the speed of sound. 
Assuming that the source is in the far field of the CMA, the direct path components in the $i$-th and $j$-th microphones are related as \vspace*{-1 mm}
\begin{equation}
X_{j}^{D}(\omega) = X_{i}^{D}(\omega)\exp(\ji \omega \tau_{ij}(\vbf_{s})) \; ,
\label{eq: direct path RDTF}
\end{equation}
where $\tau_{ij}(\vbf_{s}) = -\vbf_{s}^{\text{T}}(\m_{i} - \m_{j})/\nu$ denotes the TDOA between the $i$-th and $j$-th microphones. 
The time-difference of arrival (TDOA) is defined relative to the source DOA vector $\vbf_{\s} = [\cos(\theta_{s}), \sin(\theta_{s})]^{\text{T}}$, where $\theta_{s}$ denotes the azimuthal angle of the source, relative to the centroid of the CMA. 
The SRP function \cite{omologo1997use, dibiase2000high, dmochowski2010steered, pertila2018multichannel,dietzen2021low} is defined as \vspace*{-1 mm}
\begin{equation}
\varphi(\vbf) = \sum_{(i,j) \, : \; i>j} \int_{-\infty}^{\infty} 
{\psi}_{ij}(\omega) \exp(\ji \omega \tau_{ij}(\vbf)) 
d\omega \; ,
\label{eq: power function}\vspace*{-1 mm}
\end{equation}
with DOA vector $\vbf = [\cos(\theta), \sin(\theta)]^{\text{T}}$. 
The weighted SRP spectrum between the $i$-th and $j$-th microphones is defined as \vspace*{-1 mm}
\begin{equation}
{\psi}_{ij}(\omega) = W_{ij}(\omega) \underbrace{\EX\{Y_{i}(\omega)^{}Y_{j}^{*}(\omega)\}}_{\tilde{\psi}_{ij}(\omega)} \; ,
\label{eq: weighted SRP spectrum}\vspace*{-1 mm}
\end{equation}
with $\EX\{\cdot\}$ the expected value operator, $\{\cdot\}^{*}$ the complex conjugate operator, $\tilde{\psi}_{ij}(\omega)$ the (unweighted) SRP spectrum and $W_{ij}(\omega)$ a weighting determining the contribution of the SRP spectrum in each frequency to the integral in \eqref{eq: power function}. 

The source DOA vector is then estimated as the DOA vector maximizing the SRP function in \eqref{eq: power function}, i.e., \vspace*{-1 mm}
\begin{equation}
\hat{\vbf}_s \; = \; \argmax_{\vbf} \; \varphi(\vbf) \; .
\label{eq: DOA estimation}\vspace*{-1 mm}
\end{equation}
In this paper, we consider the commonly used PHAT weighting $W_{ij}^{\text{PHAT}}(\omega) = 1/|{\tilde{\psi}}_{ij}(\omega)|$ in \eqref{eq: weighted SRP spectrum}, yielding the SRP-PHAT spectrum $\psi_{ij}^{\text{PHAT}}(\omega)$. 
Assuming the direct path component $X_{j}^{D}(\omega)$ and the undesired component $U_{j}(\omega)$ to be uncorrelated in each microphone, the SRP-PHAT spectrum is given by \vspace*{-1 mm}
\begin{equation}
{\psi}_{ij}^{\text{PHAT}}(\omega) = \frac{\tilde{\psi}_{ij}^{D}(\omega)+\tilde{\psi}_{ij}^{U}(\omega)}{|\tilde{\psi}_{ij}^{D}(\omega)+\tilde{\psi}_{ij}^{U}(\omega)|} \; ,\vspace*{-1 mm}
\label{eq: decomposed SRP-PHAT}
\end{equation}
with the SRP spectrum of the direct path component $\tilde{\psi}_{ij}^{D}(\omega) = \EX \{ X_{i}^{D}(\omega)X_{j}^{D,*}(\omega) \}$ and the SRP spectrum of the undesired component $\tilde{\psi}_{ij}^{U}(\omega) = \EX \{ U_{i}^{}(\omega)U_{j}^{*}(\omega) \}$. 
Assuming the undesired component to be spatially uncorrelated, i.e., $\tilde{\psi}_{ij}^{U}(\omega) = 0$ for $i \neq j$, and using \eqref{eq: direct path RDTF}, the SRP-PHAT spectrum in \eqref{eq: weighted SRP spectrum} is given by ${\psi}_{ij}^{\text{PHAT}}(\omega) = \exp{(-\ji \omega \tau_{ij}(\vbf_{s}))}$, 
for which it can be easily shown that $\hat{\vbf}_{s} = \vbf_{s}$. 

Even though in practice both assumptions (i.e., uncorrelated direct path and undesired components, spatially uncorrelated undesired component) are typically not satisfied, it has been shown that the SRP-PHAT method works rather well in real noisy and reverberant environments \cite{chen2006time,zhang2008does,velasco2016proposal}. 
However, in the presence of spatially coherent noise and reverberation, the performance may suffer, especially when using CMAs, for which the spatial coherence between the microphones is typically larger than for arrays with large inter-microphone distances. 

Instead of considering alternative weightings that are more robust to noise and reverberation \cite{valin2007robust,abutalebi2011performance,athanasopoulos2016contributions,braun2015narrowband}, in this paper we propose to compute the SRP-PHAT spectrum exploiting the availability of an additional microphone at an unknown position $\m_{A}$ (see Fig. \ref{fig: Scenario}), which is spatially separated from the CMA (analysis in the following section). 
Similarly to \cite{bruemannEmic}, but now in the frequency domain instead of the time domain, we propose to compute the auxiliary microphone-based SRP-PHAT spectrum $\psi_{ij}^{\text{PHAT-A}}(\omega)$ between the $i$-th and $j$-th microphones by multiplying the SRP-PHAT spectrum ${{\psi}}_{iA}^{\text{PHAT}}(\omega)$ between the $i$-th microphone and the auxiliary microphone with the SRP-PHAT spectrum ${{\psi}}_{Aj}^{\text{PHAT}}(\omega)$ between the auxiliary microphone and the $j$-th microphone, i.e., 
\begin{equation}
\boxed{
{{\psi}}_{ij}^{\text{PHAT-A}}(\omega) = {{\psi}}_{iA}^{\text{PHAT}}(\omega){{\psi}}_{Aj}^{\text{PHAT}}(\omega)
}
\label{eq: auxiliary microphone-based SRP spectrum}
\end{equation}
Assuming the undesired component to be spatially uncorrelated, it can easily be shown that the auxiliary microphone-based SRP-PHAT spectrum is equivalent to the SRP-PHAT spectrum without the auxiliary microphone, i.e., ${\psi}_{ij}^{\text{PHAT-A}}(\omega) = \exp{(-\ji \omega \tau_{ij}(\vbf_{s}))}$.
However, for spatially coherent noise and reverberation, the SRP-PHAT spectra ${{\psi}}_{iA}^{\text{PHAT}}(\omega)$ and ${{\psi}}_{Aj}^{\text{PHAT}}(\omega)$ between the auxiliary microphone and the microphones of the CMA are assumed to be less affected by noise and reverberation than the SRP-PHAT spectrum ${{\psi}}_{ij}^{\text{PHAT}}(\omega)$ due to the spatial separation of the auxiliary microphone from the CMA. 
This will be analysed and validated in more detail in the following sections. 

\section{Analysis of Auxiliary Microphone-Based SRP-PHAT Spectrum}
\label{sec: Analysis of auxiliary microphone-based SRP-PHAT spectrum}
The main objective of this section is to theoretically analyse the influence of the undesired component on the auxiliary microphone-based SRP-PHAT spectrum in \eqref{eq: auxiliary microphone-based SRP spectrum} compared to the conventional SRP-PHAT spectrum in \eqref{eq: decomposed SRP-PHAT}. 
To this end, we define a distortion measure and analyse its contribution across frequencies for different positions of the auxiliary microphone and different power ratios between the source and the undesired component. 

Since the source is assumed to be in the far field of the CMA, we assume that $d_{i} = d_{j} = d_{c}$, with $d_{c}$ the distance between the source and the centroid of the CMA (see Fig. \ref{fig: Scenario}). 
In addition, for the analysis in this section we assume a spherically isotropic sound field \cite{elko2001spatial} for the undesired component (noise and reverberation). 
In the next section, the findings will be validated with realistic microphone signals. 
Using the aforementioned assumptions, the SRP-PHAT spectrum in \eqref{eq: decomposed SRP-PHAT} can be written as \vspace*{-1 mm}
\begin{equation}
{\psi}_{ij}^{\text{PHAT}}(\omega) = 
\frac{
\frac{\phi_{S}^{}(\omega)}{(\xi d_{c})^{2}} \exp( -\ji \omega \tau_{ij}(\vbf_{s})) + \phi_{U}(\omega) \sinc(\frac{\omega d_{ij}}{\nu}) 
}{
| \frac{\phi_{S}^{}(\omega)}{(\xi d_{c})^{2}} \exp( -\ji \omega \tau_{ij}(\vbf_{s})) + \phi_{U}(\omega) \sinc(\frac{\omega d_{ij}}{\nu}) |
} \label{eq: SRP-PHAT 1},
\end{equation}
with $\phi_S(\omega) = \EX\{|S(\omega)|^2\}$ and $\phi_{U}(\omega) = \EX\{|U(\omega)|^2\}$ the power spectral densities of the source and undesired component, respectively. 
By defining the source-to-undesired ratio (SUR) at the source position as $ \text{SUR}(\omega) = \phi_{S}(\omega) /\phi_{U}(\omega)$, the expression in \eqref{eq: SRP-PHAT 1} can be written as \vspace*{-3 mm}
\begin{equation}
{\psi}_{ij}^{\text{PHAT}}(\omega) = \frac{ 
\exp( -\ji \omega \tau_{ij}(\vbf_{s})) + \overbrace{
\textstyle
\frac{(\xi d_{c})^{2}}{\text{SUR}(\omega)} \sinc\left(\frac{\omega d_{ij}}{\nu}\right)}^{D_{ij}(\omega)}
}{
| \exp( -\ji \omega \tau_{ij}(\vbf_{s})) + \frac{(\xi d_{c})^{2}}{\text{SUR}(\omega)} \sinc\left(\frac{\omega d_{ij}}{\nu}\right) |
} \; ,\label{eq: adjusted SRP spectrum}
\end{equation}
with $D_{ij}(\omega)$ the distortion to the optimal phase shift $\exp{(-\ji \omega \tau_{ij}(\vbf_{s}))}$. 
This distortion depends on the SUR, the distance $d_{c}$ between the source and the centroid of the CMA, and the distance $d_{ij}$ between the $i$-th and $j$-th microphones. 

Similarly to \eqref{eq: adjusted SRP spectrum}, we can also rewrite the auxiliary microphone-based SRP-PHAT spectrum in \eqref{eq: auxiliary microphone-based SRP spectrum} as 
\begin{subequations}
\begin{align}
{\psi}_{ij}^{\text{PHAT-A}}(\omega) 
&= 
\frac{
 \exp( -\ji \omega \tau_{iA}(\vbf_{s})) + \frac{\xi^{2} d_{c} d_{A}}{\text{SUR}(\omega)}  \sinc(\frac{\omega d_{Ai}}{\nu}) 
}{
| \exp( -\ji \omega \tau_{iA}(\vbf_{s})) + \frac{\xi^{2} d_{c} d_{A}}{\text{SUR}(\omega)}  \sinc(\frac{\omega d_{Ai}}{\nu})  |
}
\cdot\dotsc \notag\\
&\quad\quad\quad 
\frac{
 \exp( -\ji \omega \tau_{Aj}(\vbf_{s})) + \frac{\xi^{2} d_{c} d_{A}}{\text{SUR}(\omega)}  \sinc(\frac{\omega d_{Aj}}{\nu}) 
}{
| \exp( -\ji \omega \tau_{Aj}(\vbf_{s})) + \frac{\xi^{2} d_{c} d_{A}}{\text{SUR}(\omega)}  \sinc(\frac{\omega d_{Aj}}{\nu})  |
}  ,\\
&= 
\frac{
 \exp( -\ji \omega \tau_{ij}(\vbf_{s})) + D_{ij}^{\text{A}}(\omega) 
}{
| \exp( -\ji \omega \tau_{ij}(\vbf_{s})) + D_{ij}^{\text{A}}(\omega) |
} \; ,
\label{eq: adjusted auxiliary microphone-based SRP spectrum}
\end{align}
\end{subequations}
with $d_{A}$ the distance between the source and the auxiliary microphone position and $D_{ij}^{\text{A}}(\omega)$ the auxiliary microphone-based distortion, defined in \eqref{eq: distortion}. 
\begin{figure*}[!b]
	\hrule
	\normalsize
	\setcounter{MYtempeqncnt}{\value{equation}}
	\begin{equation}
	\scalebox{0.96}{
    $D_{ij}^{\text{A}}(\omega) = 
    \frac{\xi^{2} d_{c} d_{A}}{\text{SUR}(\omega)} \left[ \sinc(\omega d_{Ai} / \nu) \exp( -\ji \omega \tau_{Aj}(\vbf_{s})) +  \sinc(\omega d_{Aj} / \nu)\exp( -\ji \omega \tau_{iA}(\vbf_{s}))\right]
    + 
    \frac{\xi^{4} d_{c}^{2} d_{A}^{2}}{\text{SUR}^{2}(\omega)} \sinc(\omega d_{Ai} / \nu) \sinc(\omega d_{Aj} / \nu)
    $
    }
    \label{eq: distortion}
	\end{equation}
	\setcounter{equation}{\value{MYtempeqncnt}+1}
\end{figure*}
It is not straightforward to analytically show under which conditions the absolute value of the auxiliary microphone-based distortion $|D_{ij}^{\text{A}}(\omega)|$ in \eqref{eq: distortion} is smaller than the absolute value of the distortion $|D_{ij}^{}(\omega)|$ in \eqref{eq: adjusted SRP spectrum}, leading to a more reliable SRP-PHAT spectrum. 
However, due to the overall decaying nature of the sinc-function and the fact that for most microphone constellations the distances $d_{Ai}$ and $d_{Aj}$ between the auxiliary microphone and the CMA are larger than the inter-microphone distances $d_{ij}$, the chances are high that the condition $| D_{ij}^{\text{A}}(\omega) |  < | D_{ij}^{}(\omega) | $ is satisfied. 
To investigate the influence of the constellations, SURs and frequencies on the distortions, we analyse the proportion of frequencies up to $\omega_{0}$ for which this condition is satisfied, i.e., \vspace{-1.2 mm}
\begin{equation}
    P \; = \; 
    \frac{1}{\omega_{0}}\int_{0}^{\omega_{0}}
    \mathcal{H}\left( \;
    | D_{ij}^{}(\omega) | - | D_{ij}^{\text{A}}(\omega) | 
    \; \right) 
    d\omega \; , \vspace{-1 mm}
    \label{eq: Percentage of frequencies}
\end{equation}
with $\mathcal{H}(x)$ the Heaviside step function, which returns the value $\mathcal{H}(x) = 0$ if $x \leq 0$ and $\mathcal{H}(x) = 1$ if $x > 0$. 
Since the PHAT weighting normalizes each frequency so that it has an equal contribution in \eqref{eq: power function}, it is assumed that when $P>0.5$, the auxiliary microphone-based SRP-PHAT function is more suitable for source localization than the SRP-PHAT function without the auxiliary microphone. 
To analyse the influence of the auxiliary microphone position on the proportion in \eqref{eq: Percentage of frequencies}, we consider an exemplary constellation, consisting of a CMA with $M=2$ microphones and inter-microphone distance $d_{12} = 5$ cm, and a source $d_{c} = 2$ m from the centroid of the CMA (see Fig. \ref{fig: Distortion Comparison}). 

To average out the influence of the microphone pair orientation, we average the proportion in \eqref{eq: Percentage of frequencies} over 18 different orientations in the x-y plane from $0^{\circ}$ to $170^{\circ}$. 
For simplicity, we assume $\text{SUR}(\omega)$ to be constant for all frequencies, i.e., $\text{SUR}(\omega) = \text{SUR}$, considering 1025 frequencies between 0 and 8 kHz. 
For different SURs, Fig. \ref{fig: Distortion Comparison} shows the influence of the auxiliary microphone position on the averaged proportion of frequencies $P_{\text{avg}}$. 
The red and blue boundaries at values $P_{\text{avg}}$ equal to 0.5 or 0.9 indicate the auxiliary microphone positions for which in 50\% or 90\% of the frequencies, respectively, the distortion of the auxiliary microphone-based SRP-PHAT spectrum is smaller than the distortion of the SRP-PHAT spectrum without the auxiliary microphone. 
It can clearly be observed that the averaged proportion $P_{\text{avg}}$ depends both on the SUR as well as on the auxiliary microphone position. 
For the auxiliary microphone to improve the reliability of the SRP-PHAT function, it should be located far enough from the CMA. 
As the SUR decreases, the size of the bounded areas notably increases. 
This means that for increasing levels of noise and reverberation, the auxiliary microphone needs to be located farther from the CMA to yield a benefit. 
It should however be noted that for $\text{SUR} \geq 10$ dB the auxiliary microphone only needs to be located 20 cm from the centroid of the CMA, whereas for $\text{SUR} \geq 0$ dB this increases to 75 cm. 
Interestingly, incorporating an auxiliary microphone can be beneficial even when it is located farther from the source than the CMA, as long as it is spatially separated from the CMA. 
Similar results are also obtained for other values of $d_{12}$ and $d_{c}$. 

\section{Experimental Validation}\label{sec: Evaluation}
To experimentally validate the findings from the previous section, in this section the DOA estimation performance of the SRP-PHAT method without and with an auxiliary microphone is compared based on simulated microphone signals. 
The performance is compared for various positions of the auxiliary microphone and different levels of noise and reverberation. \vspace{-0.2 mm}

\subsection{Scenario and Algorithm Parameters}
\label{sec: Scenario}
For the simulations, we considered a rectangular room with dimensions $6 \text{ m}\times{}6\text{ m}\times{}2.4\text{ m}$ and simulated room impulse responses using the image source method \cite{allen1979image, HabetsRIR}, assuming equal reflection coefficients for all walls. 
We considered a similar constellation as in Section \ref{sec: Analysis of auxiliary microphone-based SRP-PHAT spectrum}, now consisting of a CMA with $M=3$ microphones (to estimate the DOA without ambiguity), centered at $\m_{c} = [4,3,1.75]$ and with inter-microphone distance $d_{ij} = 5$ cm, and a source position $s = [2,3,1.75]$ m, located $d_{c} = 2$ m from the centroid of the CMA. 
We considered 601 auxiliary microphone positions on a 2D grid in the x-y plane, with an increased spatial resolution close to the CMA (see Fig. \ref{fig: Simulation Results}). 
To average out the influence of the orientation of the CMA, 12 azimuthal orientations of the CMA in the x-y plane were considered from $0^{\circ}$ to $110^{\circ}$. 
For each orientation, we considered 10 different speech signals of length 3 s, randomly selected from \cite{M-AILABS} (with equal probability for a male or female speaker), resulting in a total of 120 simulated scenarios for each auxiliary microphone position. 
To compare directly with two of the plots from Fig. \ref{fig: Distortion Comparison} where the SUR was equal to 15 dB and 0 dB, we considered two noise and reverberation conditions with similar broadband SURs. 
In the first condition, the broadband direct-to-reverberant ratio (DRR) and the broadband reverberant signal-to-noise ratio (RSNR) were both equal to -1.4 dB across the microphones of the CMA, corresponding to a broadband SUR of 14.9 dB at the source position, while in the second condition, the RSNR and the DRR were both equal to -7.2 dB across the microphones of the CMA, corresponding to a broadband SUR of 0.0 dB at the source position. 
For each condition, the reflection coefficients were determined which achieve the desired average DRR across the microphones, leading to $T_{60} = 0.18$ s for the first condition and $T_{60} = 0.32$ s for the second condition. 

The algorithms were implemented at a sampling frequency of $16$ kHz and using a short-time Fourier transform with a frame length of $512$ samples (corresponding to 32 ms), 50\% overlap between frames, and using a square-root-Hann analysis window.  
The SRP-PHAT spectra in \eqref{eq: weighted SRP spectrum} and \eqref{eq: auxiliary microphone-based SRP spectrum} were computed using recursive averaging, i.e., 
\vspace*{-1 mm}
\begin{equation}
\tilde{\psi}_{i,j}[k,l] \;\; = \;\; \lambda \tilde{\psi}_{i,j}[k,l-1] + (1-\lambda)Y_{i}Y^{*}_{j}[k,l] \; ,\vspace*{-1 mm}
\end{equation}
with frequency bin $k \in \{1,\dots,K\}$, time frame $l \in \{1,\dots,L\}$, and smoothing factor $\lambda = 0.98$ (corresponding to an averaging over approximately 0.8 s). 
The DOA estimate in \eqref{eq: DOA estimation} was obtained by first averaging the SRP-PHAT spectra over all time frames, i.e., $1/L \sum_{l=1}^{L} {\psi}_{i,j}^{\text{PHAT}}[k,l]$, and then using the averaged SRP-PHAT spectra in \eqref{eq: power function}, where the integral was approximated by a sum over discrete frequency bins $k$. 

\subsection{Comparison of DOA Estimation Performance}
To compare the DOA estimation performance of the SRP-PHAT method without and with an auxiliary microphone, we evaluated the DOA estimation error \vspace*{-1 mm}
\begin{equation}
\varepsilon_{} = \text{cos}^{-1}\left( \frac{\hat{\vbf}_{s}^{\text{T}} \vbf_{s}^{}}{||\hat{\vbf}_{s}^{}||_2\cdot||\vbf_{s}^{}||_2} \right) \; ,\vspace*{-1 mm}
\end{equation}
averaged over the 120 simulated scenarios. 

\begin{figure}[t!]
    \centering
    \includegraphics[width=\linewidth]{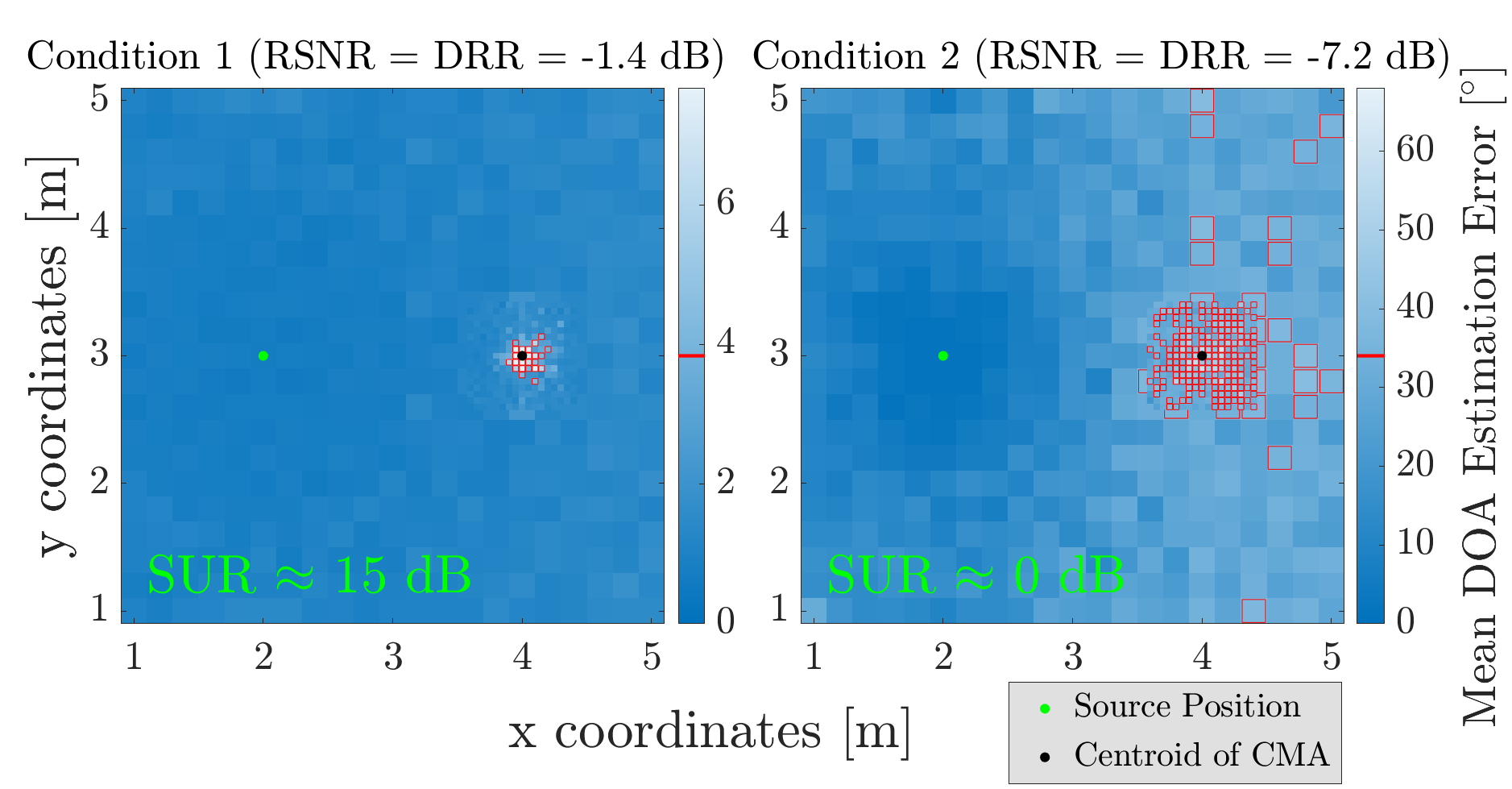}
    \caption{Mean DOA estimation error of the auxiliary microphone-based SRP-PHAT function compared to the SRP-PHAT function without the auxiliary microphone, for different auxiliary microphone positions in a 2D plane and two different noise and reverberation conditions.\vspace*{-2 mm} }
    \label{fig: Simulation Results}
\end{figure}

For all considered auxiliary microphone positions, Fig. \ref{fig: Simulation Results} illustrates the mean DOA estimation error, for both noise and reverberation conditions. 
\begin{figure*}[t!]
    \centering
    \includegraphics[width=1\linewidth]{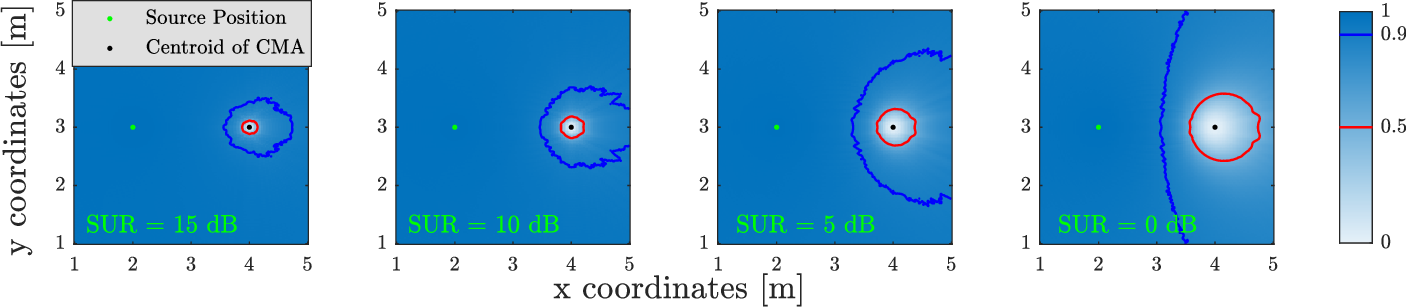}
    \caption{Averaged proportion of frequencies for which the distortion of the auxiliary microphone-based SRP-PHAT spectrum is smaller than the distortion of the SRP-PHAT spectrum without the auxiliary microphone, where each point in the 2D plane corresponds to a different position of the auxiliary microphone.}\vspace*{-2 mm}
    \label{fig: Distortion Comparison}
\end{figure*}
The mean DOA estimation error for the SRP-PHAT method without an auxiliary microphone was equal to $3.8^{\circ}$ in the first condition (RSNR $=$ -1.4 dB, DRR $=$ -1.4 dB), and $34.0^{\circ}$ in the second condition (RSNR $=$ -7.2 dB, DRR $=$ -7.2 dB). 
It can clearly be seen that incorporating an auxiliary microphone in the SRP-PHAT method generally results in a reduced mean DOA estimation error
, especially for larger distances between the auxiliary microphone and the CMA. 
This aligns well with our findings from Section \ref{sec: Analysis of auxiliary microphone-based SRP-PHAT spectrum}. 

The red-bordered pixels in Fig. \ref{fig: Simulation Results} indicate positions of the auxiliary microphone for which incorporating the auxiliary microphone did not reduce the mean DOA estimation error. 
First, it can be seen that in the vicinity of the CMA the spread of red-bordered pixels is larger for the more challenging noise and reverberation condition. 
This corresponds to the area bounded by the red line in Fig. \ref{fig: Distortion Comparison}, which increases as the SUR decreases. 
Second, it can be seen that certain auxiliary microphone positions in the vicinity of walls exhibit a slightly larger mean DOA estimation error when the auxiliary microphone is used. 
This can be attributed to the limitations of the assumed model in Section \ref{sec: Analysis of auxiliary microphone-based SRP-PHAT spectrum}, which assumed a spherically isotropic field for the reverberation without considering early reflections from the walls. 

\section{Conclusions}\label{sec: Conclusions}
In this paper, we have proposed a method to improve SRP-PHAT-based DOA estimation in CMAs, where noise and reverberation with a high spatial coherence may negatively affect the reliability of the SRP-PHAT spectra. 
Assuming the availability of a spatially separated auxiliary microphone, we proposed to compute the SRP-PHAT spectra in a different way, using the SRP-PHAT spectra between the auxiliary microphone and the microphones of the CMA. 
For different SURs, we showed how far the auxiliary microphone needs to be spatially separated from the CMA for the SRP-PHAT spectra to be estimated more reliably with the auxiliary microphone than without. 
Furthermore, we validated these findings with simulation results based on simulated microphone signals for many different auxiliary microphone positions and two different noise and reverberation conditions, showing the beneficial impact of incorporating an auxiliary microphone for DOA estimation. 

\bibliographystyle{IEEEtran}
\balance
\bibliography{ms}

\end{document}